# The Availability of Logical Operation Induced by Dichotomous Noise for a Nonlinear Bistable System

Yong Xu[1], Xiaoqin Jin, Huiqing Zhang, Tingting Yang

Department of Applied Mathematics

Northwestern Polytechnical University, Xi'an, 710072, China

**ABSTRACT:** Instead of a continuous system driven by Gaussian white noise, logical stochastic resonance will be investigated in a nonlinear bistable system with two thresholds driven by dichotomous noise, which shows a phenomenon different from Gaussian white noise. We can realize two parallel logical operations by simply adjusting the values of these two thresholds. Besides, to quantify the reliability of obtaining the correct logic output, we numerically calculate the success probability, and effects of dichotomous noise on the success probability are observed, these observations show that the reliability of realizing logical operation in the bistable system can be improved through optimizing parameters of dichotomous noise.

## 1. INTRODUCTION

Over the past few decades, noise has always been perceived as a disturbance in changing the desired output signals. However, noise has been verified to play a constructive role in stochastic resonance and can enhance the output signal of a nonlinear system. Benzi and Nicolis [1] originally proposed stochastic resonance (SR) to deal with the problems about the regularly recurrent ice ages. Gammaitoni [2] suggested SR to be an example to illustrate how to understand the cooperative behavior between noise and nonlinear dynamics. Besides, stochastic resonance has attracted immense attentions and has been applied in many fields, such as biology [3], chemistry [4], electronic circuits [5, 6], lasers [7], and so on.

[1] Corresponding author. E-mail:hsux3@nwpu.edu.cn.
 Tel. /fax numbers: 86-29-88431637.

Considering such a fact that electronic components keep on shrinking in size, how to minimize the influence of noise in computational devices and electronic circuits has been widely studied recently. However, moderate noise can not degrade computation but facilitating computation, which becomes more significant to understand the cooperative between noise and nonlinearity in the design of computational devices.

By exploring the interplay between noise and nonlinearity, Murali et al has designed noisy logic gate [8], which is also named as logical stochastic resonance (LSR) and used to logic computation of SR. Although LSR is a relatively new idea, variety of former studies focused greatly on its properties and applications. For instance, Murali and Rajamohamed [8] discussed the response of a simple threshold detector to input signals. Singh and Sinha [9] verified the reliability of LSR in a polarization bistable laser and processed two parallel logic gates in the laser system. In order to observe and implement LSR in electronic field, Murali and Sinha [10] presented an electronic circuit with CMOS transistors, and suggested that the fundamental logic operations NOR or NAND were achieved in an optimal wide band of noise and one logic operation can be switched into another. Zhang and Song [11] showed the effect of LSR in an asymmetric bistable model in the presence of $1/f^{\beta}$ noise and obtained a kind of SR-like effect controlled by noise exponent. Moreover, in the absence of noise but a period random fluctuation, a phenomenon analogous to LSR can be achieved in an optimal range of frequency and amplitude of fluctuation [12]. Furthermore, the effect of LSR has been extended to a synthetic gene network, in which Dari and Kia [13] have proved the possibility of obtaining the biological logic gate in a gene regulatory network (GRN). The LSR behavior has been examined in physical systems, extending from electrical and nanomechanical [14] to optical systems [15], chemical and biological scenarios [16]. Besides, the confined system of sharp geometric constrictions [17, 18] may be available to the logic gate operation due to that this system is a three-dimensional equation, which can be reduced to a one-dimensional equation by the Fick-Jacobs approximation method [19]. Since the effects of the confinement are studied via both an entropic potential and space dependent diffusivity in this reduced equation, so the effective bilobal confinement with sharp or smooth wall is regarded as a bistable system. And the asymmetry of this bistable potentials changes with the shape of the confinement changing. However, some problems may be encountered in the realization of logic gate in this confined system. For instance, the bistability of the confinement may not be enough to obtain the logic gate in this case, and some external forces need to be added. Based on these descriptions, the LSR may be realized in the system of sharp geometric constrictions in the future.

Up to now, most of the existing work studied the effect of LSR in bistable systems based on the assumption of Gaussian white noise. But there is no study on the achievement of logic gate in a bistable system induced by dichotomous noise, which is also named random telegraphic noise jumping between two values, and have been considered in stochastic resonance [16, 23]. Then a question naturally arises: Does there exist a possibility to exhibit the LSR phenomenon in a bistable system induced by a dichotomous noise?

Our motivation of choosing dichotomous noise instead of Gaussian white noise to give rise to LSR results from that: the dichotomous noise is superior to Gaussian white noise, because the dichotomous noise can approach to both Gaussian white noise and white shot noise under well-defined limiting procedures [24, 25]. Additionally, the dichotomous or telegraphic noise is a better representation of real noise than the widely used Gaussian white noise [25], in many cases, such as metal [26], bipolar transistor [27], nanometer devices [28] and so on. And the dichotomous noise is presented in the form of electric resistance [26], electric conductance [29] and voltage [28]. Thus, studying the dichotomous noise has more important practical significance.

In this paper, the analysis of LSR induced by the dichotomous noise in a bistable system shall be the main purpose of our upcoming work. The framework of this paper are as follows: In Sec 2 we mainly study the effect of LSR for a bistable system induced by dichotomous noise with thresholds $(x_l, x_u) = (-1.5, 0.5)$, introduce the calculation of the escape rate and present the influences of the noise asymmetry, correlation time and noise intensity on the success probability. Sec 3 demonstrates that another logic operation can be obtained by changing the thresholds $(x_l, x_u) = (-0.5, 1.5)$ of the bistable system with any fixed correlation time. Moreover, the influence of the noise correlation time on the success probability has been investigated, and the dichotomous noise is superior to Gaussian white noise in enhancing the logical stochastic resonance phenomenon. Finally Sec 4 offers our conclusions and discussions.

## 2. THE RELIABILITY OF LSR FOR BISTABLE SYSTEM WITH THRESHOLDS $(x_l, x_u) = (-1.5, 0.5)$

### 2.1. The System Model and Escape Rate

Consider a nonlinear dynamic system described by the following stochastic differential equation [10]:

$$\dot{x} = f(x) + I(t) + DQ(t)$$
$$= -ax + bg(x) + I(t) + DQ(t), \qquad (1)$$

where

$$\dot{x} = \frac{dx(t)}{dt}, \quad f(x) = -\dot{U}(x) = -\frac{dU(x)}{dx} = -ax + bg(x),$$

$U(x)$ is the double well potential function with respect to $x$. $a$, $b$ are constant coefficients and $g(x)$ is a piecewise linear function as below:

$$g(x) = \begin{cases} x_l, & x < x_l, \\ x, & x_l \le x \le x_u, \\ x_u, & x > x_u, \end{cases} \quad (2)$$

in which $x_l$ denotes the lower thresholds and $x_u$ denotes the upper one to control the depth of the two side wells. It is possible to arrive at the desired two values depending on appropriate choice, here we take

$$x_l = -1.5, \quad x_u = 0.5, \quad a = 1, \quad b = 2. \quad (3)$$

So we can get two minimum values and a maximum value from the potential function $U(x)$:

$$x_{in} = -3, \quad x_{out} = 1, \quad x_{top} = 0. \quad (4)$$

Besides, in system (1), $Q(t)$ is an asymmetric dichotomous noise [20, 23], which rotates between two state values $\Delta_1 < 0$ and $\Delta_2 > 0$. The transition rates of $Q(t)$ from $\Delta_2$ to $\Delta_1$ and vice versa are denoted by $\alpha$ and $\beta$, respectively. This two-step process can be described with a probability loss-gain equation [20-24]

$$\frac{d}{dt} P(\Delta_1, t | x, t_0) = -\alpha P(\Delta_1, t | x, t_0) + \beta P(\Delta_2, t | x, t_0), \quad (5)$$

$$\frac{d}{dt} P(\Delta_2, t | x, t_0) = \alpha P(\Delta_1, t | x, t_0) - \beta P(\Delta_2, t | x, t_0), \quad (6)$$

where $P(\Delta_1, t | x, t_0)$ is a conditional probability, which represents that $Q(t)$ takes value $\Delta_1$ at some time given that it takes value $x$ at earlier time $t_0$. Similarly, $P(\Delta_2, t | x, t_0)$ can be defined.

The sum of $P(\Delta_1, t | x, t_0)$ and $P(\Delta_2, t | x, t_0)$ demands

$$P(\Delta_1, t | x, t_0) + P(\Delta_2, t | x, t_0) = 1, \quad (7)$$

at $t = t_0$, the initial condition are given by

$$P(x',t|x,t_0) = \delta_{x'x}, \quad x, x' = \Delta_1 \text{ or } \Delta_2, \tag{8}$$

then we can obtain the conditional probabilities $P(\Delta_1,t|x,t_0)$ and $P(\Delta_2,t|x,t_0)$ for time $t$ by solving Eqs. (5) and (6) subjected to Eqs. (7) and (8):

$$P(\Delta_1,t|x,t_0) = \alpha/(\alpha+\beta) + \left((\alpha\delta_{\Delta_1 x} - \beta\delta_{\Delta_2 x})/(\alpha+\beta)\right)\exp(-(\alpha+\beta)(t-t_0)), \tag{9}$$

$$P(\Delta_2,t|x,t_0) = \alpha/(\alpha+\beta) - \left((\alpha\delta_{\Delta_1 x} - \beta\delta_{\Delta_2 x})/(\alpha+\beta)\right)\exp(-(\alpha+\beta)(t-t_0)). \tag{10}$$

According to Eqs. (9) and (10), the steady-state probabilities $P_{st}(\pm\Delta)$ are given by [30]

$$P_{st}(\Delta_2) = P(\Delta_2,\infty|x,t_0) = \beta/\gamma, \tag{11}$$

$$P_{st}(\Delta_2) = P(\Delta_2,\infty|x,t_0) = \beta/\gamma, \tag{12}$$

$\gamma = \alpha + \beta$. In the following we will consider the statistical properties of the asymmetric dichotomous noise. The noise intensity $D$ and correlation time $\tau$ can be defined as:

$$D = \int_{-\infty}^{+\infty} \langle Q(t)Q(0)\rangle dt/2 = \tau|\Delta_1|\Delta_2. \tag{13}$$

Then the mean and correlation function of the asymmetric dichotomous noise are given as follows [30]:

$$\langle Q(t)\rangle = (\Delta_1\alpha + \Delta_2\beta)/(\alpha+\beta), \tag{14}$$

$$\langle Q(t_1)Q(t_2)\rangle = D e^{-|t_1-t_2|/\tau}/\tau, \tag{15}$$

$$\tau = 1/(\alpha+\beta) = 1/\gamma. \tag{16}$$

In Eq. (15), $t_1$ is a time variable, $t_2 = t_1 + Dt$, $Dt$ is a time step. Apparently, from (15), we can immediately obtain that the dichotomous noise is exponentially correlated. Moreover, we introduce the asymmetry parameter $A$ for the dichotomous noise [31]:

$$A = (\Delta_2 - |\Delta_1|)/(\Delta_2 + |\Delta_1|). \tag{17}$$

With these above preparations, we now use the method of acceptance-rejection to describe the generation of a dichotomous noise [20, 23-24]. Let the particle initially takes the value $\Delta_1$ at time $t$, namely assuming $x_n = \Delta_1$. We determine the value of $x_{n+1}$ at next time $t_1 = t + Dt$ in the following way.

Firstly, from Eq. (9), we can give the conditional probability

$$P(\Delta_1, t_1 | \Delta_1, t) = \beta/(\alpha + \beta) - (\beta/(\alpha + \beta))\exp(-(\alpha + \beta)Dt), \tag{18}$$

then we generate a uniformly distributed random number R between [0, 1]. With this number compared against $P(\Delta_1, t_1 | \Delta_1, t)$, if $P(\Delta_1, t_1 | \Delta_1, t) > R$, then $x_{n+1}$ to be $\Delta_1$, otherwise $x_{n+1} = \Delta_2$. If $x_{n+1} = \Delta_2$ at $t_1$, we determine the value of $x_{n+2}$ at next time $t_2 = t_1 + Dt$ by calculating the probability $P(\Delta_1, t_2 | \Delta_2, t_1)$. $P(\Delta_1, t_2 | \Delta_2, t_1)$ is given by Eq. (10):

$$P(\Delta_1, t_2 | \Delta_2, t_1) = \beta/(\alpha + \beta) - (\beta/(\alpha + \beta))\exp(-(\alpha + \beta)Dt), \tag{19}$$

if $x_{n+1} = \Delta_2$ at $t_1$, we compare $P(\Delta_1, t_2 | \Delta_2, t_1)$ against another uniformly distributed random number $R_1$ between [0, 1]. If $P(\Delta_1, t_2 | \Delta_2, t_1) > R_1$, then $x_{n+2} = \Delta_1$, else $x_{n+2} = \Delta_2$. Repeating the procedure, a sequence of $Q(t)$ switching between $\Delta_1$ and $\Delta_2$ can be generated (see Fig. 1).

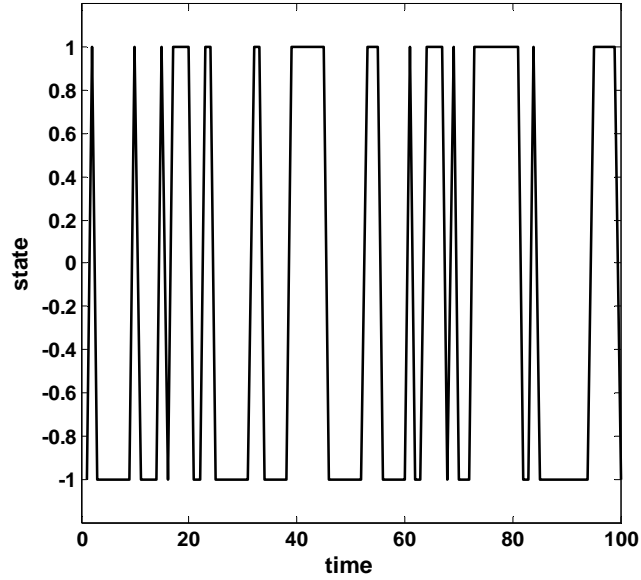

Fig.1. The dichotomous noise $Q(t)$ vs. time $t$, with $\Delta_1 = -1$, $\Delta_2 = 1$, $D = 0.01$, $Dt = 0.01$.

Besides, the master equation corresponding to system (1) for the probabilities $P(x, \Delta_1, t)$ and $P(x, \Delta_2, t)$ reads [30]:

$$\frac{\partial}{\partial t} p(x, \Delta_1, t) = -\frac{\partial}{\partial x}\{[-ax + bg(x) + I + \Delta_1] p(x, \Delta_1, t)\} - \beta p(x, \Delta_1, t) + \alpha p(x, \Delta_2, t), \quad (20)$$

$$\frac{\partial}{\partial t} p(x, \Delta_2, t) = -\frac{\partial}{\partial x}\{[-ax + bg(x) + I + \Delta_2] p(x, \Delta_2, t)\} - \alpha p(x, \Delta_2, t) + \beta p(x, \Delta_1, t), \quad (21)$$

the accurate steady-state density function $p(x)$ of the bistable system (1) has been studied and can be written as [30]:

$$p(x) = N / ([-ax + bg(x) + I + \Delta_1][-ax + bg(x) + I + \Delta_2])$$

$$\cdot \exp\left\{-\gamma \int_x ([-ax + bg(x) + I] / ([-ax + bg(x) + I + \Delta_1][-ax + bg(x) + I + \Delta_2])) dx\right\}, \quad (22)$$

where $N$ denotes a normalization constant.

Finally, the forward escape rate $x_{in} \to x_{out}$ follows from (22) can be defined as:

$$k_f = 1 \bigg/ \int_{x_{in}}^{x_{out}} p(x)\,dx. \qquad (23)$$

Additionally, by using the method of steepest descent, the forward escape rate can be evaluated [31]

$$k_f = \left(\left(-(f(x_{in})+I)'(f(x_{top})+I)'\right)^{1/2} \exp\{-\Delta\phi/\tau\}\right) \bigg/ \left(2\pi\left(1+\tau(f(x_{top})+I)'\right)\right), (24)$$

in which

$$\Delta\phi = \int_{x_{in}}^{x_{top}} \left((f(x)+I) \big/ \left((f(x)+I+\Delta_1)(f(x)+I+\Delta_2)\right)\right) dx. \qquad (25)$$

Next the effects of the noise intensity and the input signal on the forward escape rate are illustrated in Fig. 2.

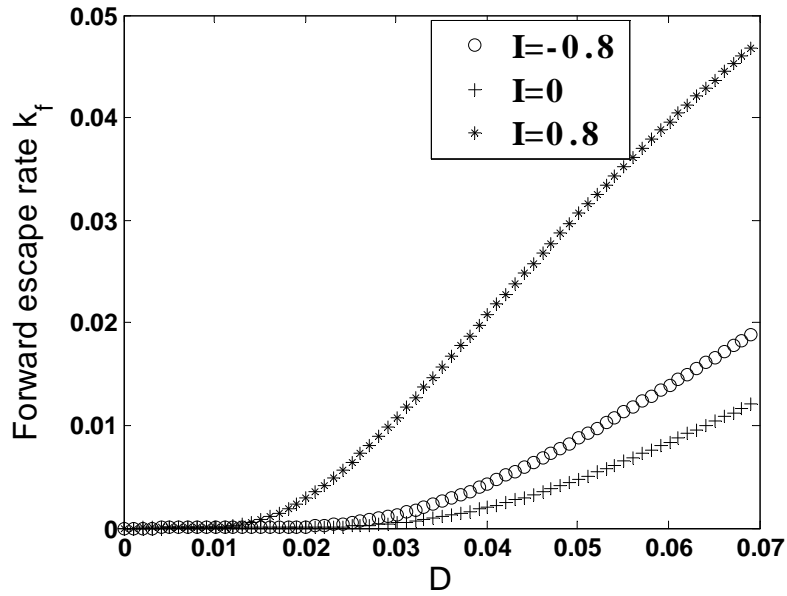

Fig.2. The forward escape rate $k_f$ vs. noise intensity with different input signals $I=-0.8$ (circle line), $I=0$ (plus line), $I=0.8$ (star line) for the model (1) with system parameters (3). With fixed dichotomous noise $\Delta_1=-1$, $\Delta_2=1$ and with a time step $Dt=0.01$.

In a completely similar way, the backward escape rate $k_b$ can also be calculated. Then the backward escape rate versus the noise intensity with different input signal is illustrated in Fig. 3.

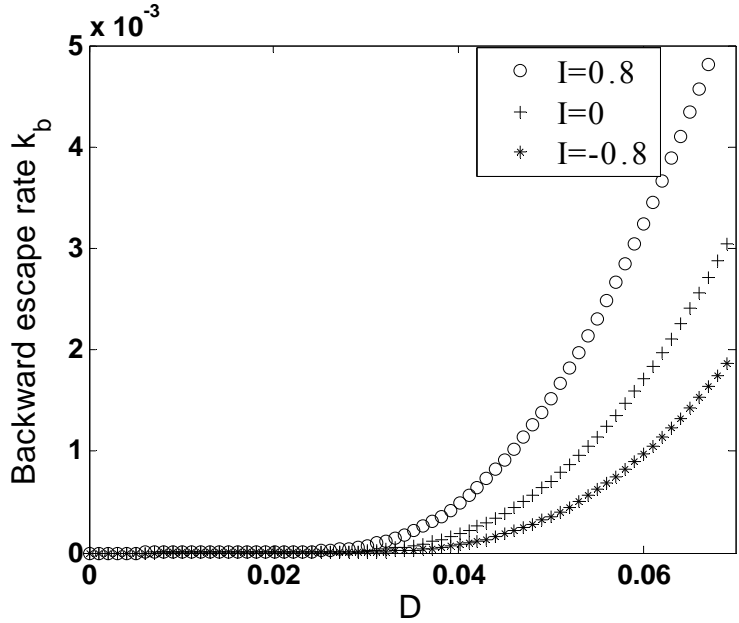

Fig.3. The backward escape rate $k_b$ vs. noise intensity with different input signals $I=-0.8$ (star line), $I=0$ (plus line), $I=0.8$ (circle line) for model (1) with system parameters (3). With fixed dichotomous noise $\Delta_1=-1$, $\Delta_2=1$ and with a time step $Dt=0.01$.

**2.2. The LSR with Fixed Noise Asymmetry**

With fixed noise asymmetry or fixed dichotomous noise values, we can achieve the system response corresponding to the logical inputs in system (1) with thresholds $(x_l, x_u)=(-1.5, 0.5)$. Here the system is driven by the summed signal $I(t)=I_1(t)+I_2(t)$, where $I_1(t)$ and $I_2(t)$ are the two logic inputs.

Without loss of generality, we set the inputs $I_i (i=1,2)$ to take a value 0.4 for the logic input 1, and value -0.4 for the logic input 0. Hence four sets of binary inputs $(I_1, I_2): (0,0), (0,1), (1,0), (1,1)$ are produced. As the sets $(0,1)$ and $(1,0)$ lead to the same value $I = I_1 + I_2$, so these four distinct sets of inputs reduce to three distinct input signals $I$ [10].

According to the well where the state $x$ is in, the logical output of the system is determined. More specifically, we consider the output to be a logical 1 if $x$ is in the right well and 0 otherwise. Thus the logical operation from system (1) can be checked by the truth table of basic logic relations.

| Input set $(I_1, I_2)$ | OR | NOR | AND | NAND |
|---|---|---|---|---|
| $(0,0)$ | 0 | 1 | 0 | 1 |
| $(1,0)/(0,1)$ | 1 | 0 | 0 | 1 |
| $(1,1)$ | 1 | 0 | 1 | 0 |

**TABLE1**. Relationship between two logic inputs and the logic output for the fundamental AND, NAND, OR, and NOR logic behaviors.

The logic inputs and the system outputs with different noise intensities of the dichotomous noise are displayed in Fig. 4.

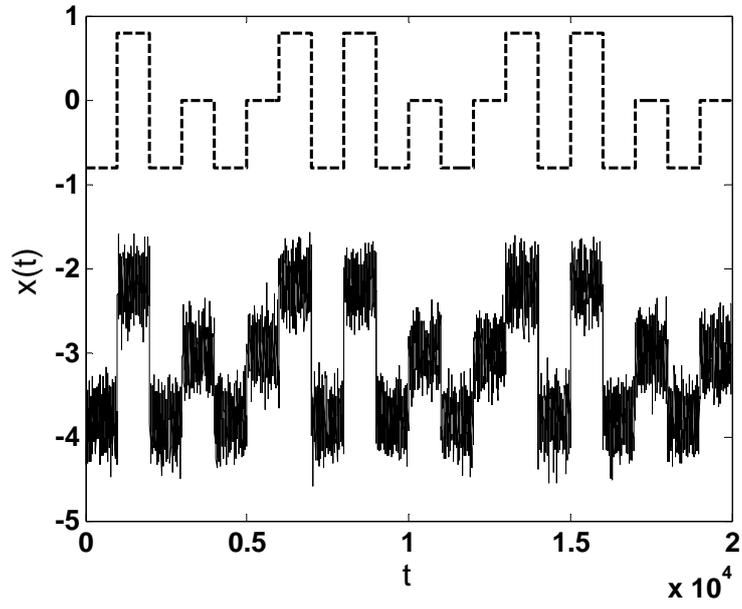

(a)

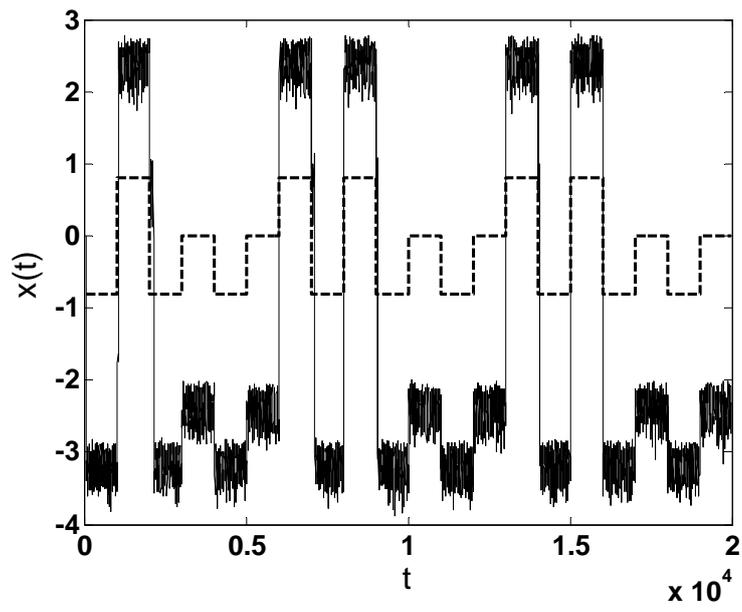

(b)

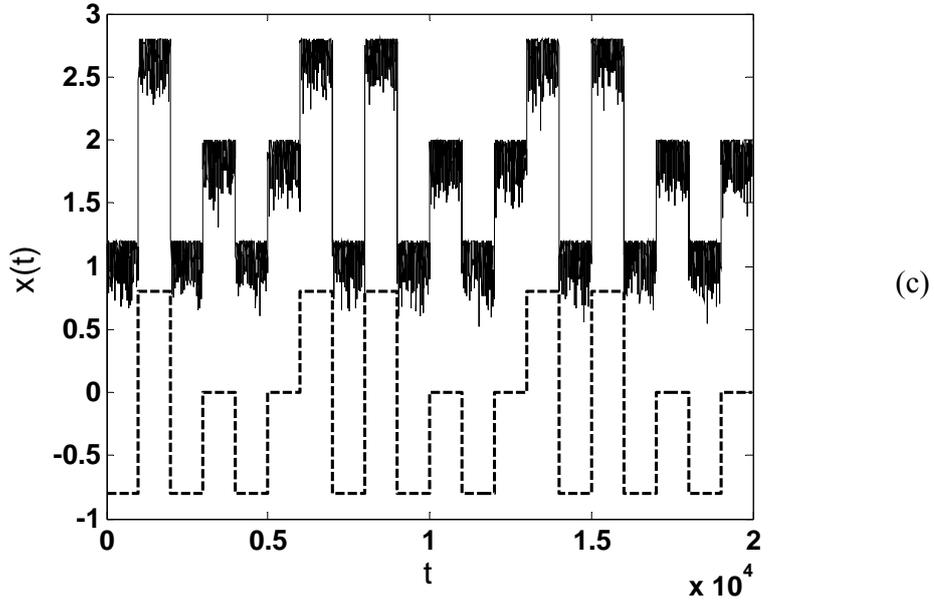

Fig.4. From top to bottom, panels 1, 2, and 3 depicts the logic input streams $I(t)$ (dash line) and output streams $x(t)$ (solid line) for three different noise intensities D: (a) D=0.0003 (b) D=0.02 (c) D=0.07. The system response of the AND gate subjected to dichotomous noise is calculated with system parameters (3), dichotomous noise values $\Delta_1 = -1$. $\Delta_2 = 1$. $I = -0.8, \ 0, 0.8$, and a fixed time step $Dt = 0.01$.

Evidently, as can be observed from Fig. 4, with fixed dichotomous noise values, the logic gate AND can be obtained. For smaller noise level, the particles fall in the left well and in the right well for larger noise. Only for medium noise intensity, the system output falls in correct well and yields an obvious logical AND. It is worth highlighting that this phenomenon has never occurred in systems subjected to Gaussian white noise. For a bistable system induced by Gaussian white noise, the system states fall in one well with fixed smaller noise intensity, whereas, the system states skip between two wells when the noise intensity is larger. And in a suitable noise level, the system obtain correct logic gate.

As to the explanation of this phenomenon, we can be traced back to Fig. 2 and Fig. 3, from which we can observe that the forward and backward escape rates exponentially increase with the increase of the noise intensity. Specifically, for $I = -0.8$ and $I = 0$, both the forward and backward escape rate are close to 0 when $D <= 0.03$, so the particles at the beginning in the left well cannot hop over the potential well but stably fall in the left well (see Fig. 4 (a), (b)). When $D > 0.03$, with the increase of the noise intensity, the forward escape rate increases quickly, which leads the particles in the left well to hop over the potential barrier and fall in the right well. Although the backward escape rate increases, but it is still too small, so the particles can not jump out from the right well (see Fig. 4 (c)). As for $I = 0.8$, the forward escape rate is close to 0 when $D <= 0.01$, thus the particles stably fall in the left well in Fig. 4 (a). When $D > 0.01$, the forward escape rate increases quickly with the increasing noise intensity, consequently, the particles can skip over the potential barrier and simultaneously fall in the right well, since the backward escape rate is close to 0, which results in that the particles would always stay in the right well (see Fig. 4 (b), (c)).

**2.3. The Influence of Asymmetry Parameter of Dichotomous Noise on LSR**

A measurement of LSR is the success probability $P(\text{logic})$, given by [10]:

$$P(\text{logic}) = (\text{the number of correct logic outputs}) / (\text{the total number of runs}). \quad (26)$$

In order to quantify the reliability of obtaining the given logic output, we numerically calculate the success probability $P(\text{logic})$, which describes the possibility of obtaining the desired logic output for different input sets. As expressed in Eq. (26), the probability $P(\text{logic})$ is the ratio of the number of correct logic outputs to the total number of runs. For each run, four input sets $(0,0)$, $(0,1)$, $(1,0)$, $(1,1)$ are presented in a random order and the run is deemed as a success only when the logic outputs obtained from $x$ matches the logic outputs in the truth table for all four input sets. The nonlinear system always yields correct logic output when $P(\text{logic})$ is close to 1.

Besides, since different logic inputs can be mapped to a binary 0/1 logical output, which is determined by the well where the system state is in, so for a nonlinear bistable system by special parameter settings, the logic gates OR and AND can be obtained. For logical OR, the system driven by input signals $(0,1)/(1,0)$ and $(1,1)$, goes to the right well and $(0,0)$ results in the system being in the left well; whereas for logical AND, the system driven by input signals $(0,0)$ and $(0,1)/(1,0)$, goes to the left well and input $(1,1)$ sends the system to the right well [32].

Next we study the LSR effect induced by changing the asymmetry parameter A (fixed dichotomous noise values). In the presence of asymmetry dichotomous noise for system (1) with parameters (3), the fundamental logical AND is realized in an optimal wide range of noise intensity. For three different asymmetry parameters of the dichotomous noise, the success probability $P(\text{AND})$ for the bistable system (Eq. (1)) under different noise intensities is figured out in Fig. 5.

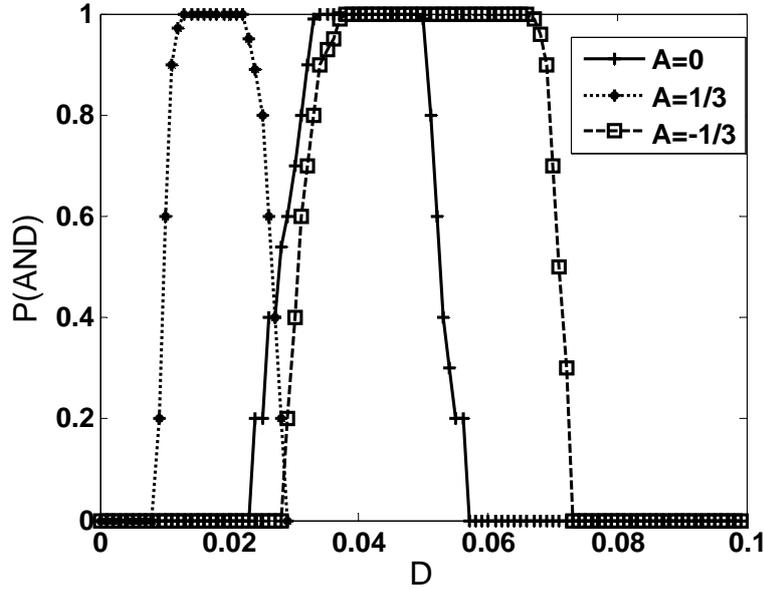

Fig.5. The success probability of obtaining the logical AND versus noise intensity $D$ with different asymmetry parameters: $A = 1/3$ (dotted-star line), $A = 0$ (solid-plus line), $A = -1/3$ (dash-rectangle line). With system parameters (3) and fixed time step Dt=0.01.

This figure shows that the success probability of the output P(AND) varies non-monotonically with the increasing noise intensity $D$. When the noise intensity is small, the success probability P(AND) increases with the increasing noise intensity. As the noise intensity increases, the success probability P(AND) reaches to the maximum and the success probability closes to 1. And with the further increase of the noise intensity, the success probability decreases and tends to zero. The fundamental logic gate AND is achieved only in a reasonably wide range of moderate noise. That is to say, in an optimal band of moderate noise, the success probability is approximately 1.i.e. $P(\text{logic}) \approx 1$. This can be understood easily, when fixed the asymmetry parameter A (fixed dichotomous noise values), the system outputs would fall in one well with smaller or larger noise intensity and fail to realize correct logic gate, so the success probability is small, whereas for appropriate noise intensity, the system outputs and inputs are in according with the logic truth table and obtained logic gate AND, so $\text{P(logic)} \approx 1$. Besides, with the decreasing $A$ value, the inverse $U$ shape curves move to right and become wider. This is due to that: the dichotomous noise value $\Delta_2$ is becoming smaller and $|\Delta_1|$ is becoming larger as the asymmetry parameter $A$ decreases, which can be known from Eq. (17). Since smaller $|\Delta_1|$ value or larger value $\Delta_2$ can lead the system states locating initially in one well to skip more easily into the other one with a fixed noise intensity. In addition, from Eq. (13), the noise intensity $D$ is directly proportional to the correlation time $\tau$, and smaller $\tau$ value results in faster skipping between two noise states, thus leading the system states to switch between two wells. So according to the above descriptions, larger asymmetry parameter $A$ and smaller noise intensity can contribute the system to achieve the correct logic gate. Therefore, the optimal noise intensity range corresponding to $\text{P(logic)} \approx 1$ becomes smaller and narrower with the asymmetry parameter $A$ increases.

## 2.4. The Influence of Noise Correlation Time on LSR

At fixed asymmetric dichotomous noise values, the logic AND gate has been obtained in the above section for the system thresholds $(x_l, x_u) = (-1.5, 0.5)$. And if the correlation time of the dichotomous noise is fixed, then the logic AND can be obtained in system (1) and the effect of the correlation time $\tau$ on LSR has been studied in Fig. 6.

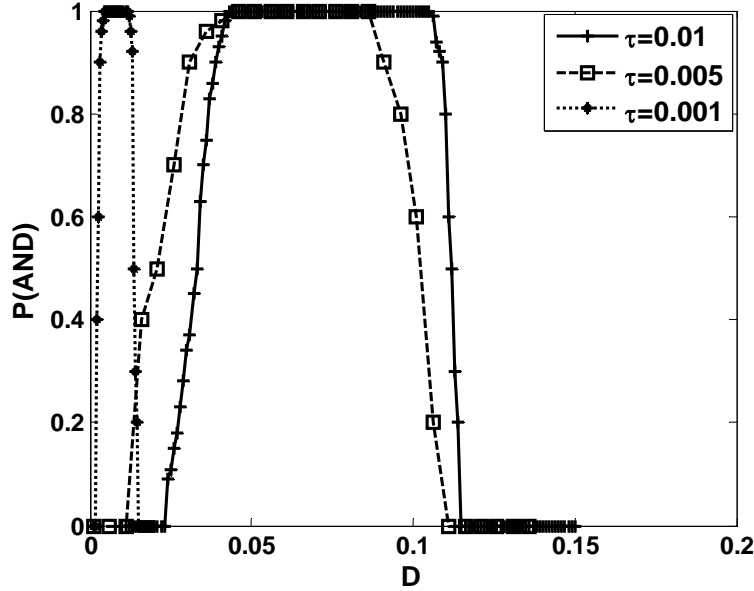

Fig.6. The success probability of obtaining the logical AND versus noise intensity $D$ with three different correlation time $\tau = 0.001$ (dotted-star line), $\tau = 0.005$ (dash-rectangle line), $\tau = 0.01$ (solid-plus line) with fixed time step Dt=0.01 and system parameters $a = 1$, $b = 2$, $x_l = -1.5$, $x_u = 0.5$, $A = 0$.

Apparently, as can be observed from Fig. 6 that with the increasing noise intensity, the success probability of obtaining the logical AND first increases, reaches to an plateau where $P(\text{logic}) \approx 1$, and then decreases at any fixed $\tau$ value. And the phenomenon of LSR occurs only in an optimal range of noise level. Similar to the explanation of Fig. 4, with fixed correlation time, the system states fall in one well and cannot realize any correct logic operation both for smaller and larger noise intensity, so the success probability is close to 0. However, for suitable noise intensity, the system can obtain the logic gate AND, so the success probability is approach to 1. Moreover, the curve of success probability P(AND) moves to right, simultaneously the noise range corresponding to the resonance peak becomes larger with the increasing correlation time $\tau$. This can be easily understood from the fact that smaller noise intensity and smaller correlation time can result in a more quickly switching between two noise values, thus leading to the system states skip between two wells more easily. Therefore, the optimal noise values to realize the peak of success probability are smaller for a fixed smaller correlation time than a larger correlation time. Namely, the curve of success probability P(AND) moves to right as the correlation time decreases.

## 2.5. The Influence of Noise Intensity on LSR

If we fix the noise intensity D in the system (1) with parameters $a=1$, $b=2$, $x_l=-1.5$, $x_u=0.5$, $A=0$, the logical AND can be obtained. The success probability of obtaining the logical AND as a function of correlation time is plotted in Fig. 7.

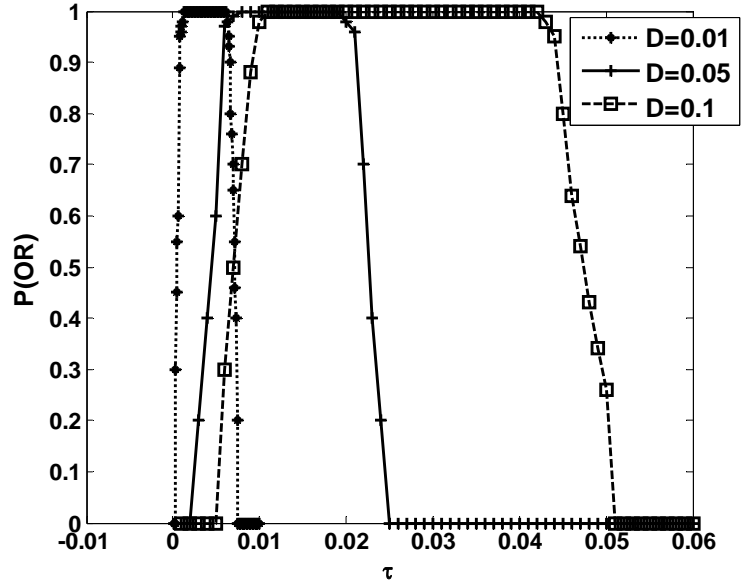

Fig.7. The probability in logical AND vs. correlation time of dichotomous noise for several values of noise intensity: D=0.01 (dotted-star line), D=0.05 (dotted-plus line), D=0.1 (dash-rectangle line) for system (1) with $a=1$, $b=2$, $x_l=-1.5$, $x_u=0.5$, $A=0$ and fixed time step Dt=0.01.

On one hand, this figure shows that with the increasing of correlation time $\tau$, the probability P(AND) firstly increase, reaches a maximum and then decreases, and the reliability of LSR can be obtained in a range of moderate $\tau$ value. This is because that smaller or larger correlation time results in the system states a too quickly or too slowly jumping between two wells, only suitable correlation time can lead the system states to jump regularly and obtain correct logic operation, hence the success probability approaches to 1 only in a moderate range of $\tau$ value. On the other hand, as the noise intensity D increases, the curve of the probability P(AND) moves to right and the correlation time range for $P(AND) \approx 1$ becomes larger. As to explain this, we make such an analysis, that is, smaller correlation time or smaller noise intensity can contribute the system states to switch between two wells, so correct logic gate can be achieved under a relatively small suitable correlation time range and fixed smaller noise intensity. Namely, the optimal range of correlation time corresponding to $P(AND) \approx 1$ becomes larger as the noise intensity decreases.

## 3. THE RELIABILITY OF LSR FOR BISTABLE SYSTEM WITH THRESHOLDS $(x_l, x_u) = (-0.5, 1.5)$

In a completely similar way to system (1) with thresholds $(x_l, x_u) = (-1.5, 0.5)$, considering that changing the thresholds $(x_l, x_u)$ changes the position, depths and asymmetry of the potential wells, so when we set the system thresholds $(x_l, x_u) = (-0.5, 1.5)$, the logical OR can be obtained with a fixed correlation time $\tau$. The success probability of obtaining the logical OR as a function of noise intensity for different $\tau$ values in Fig. 8.

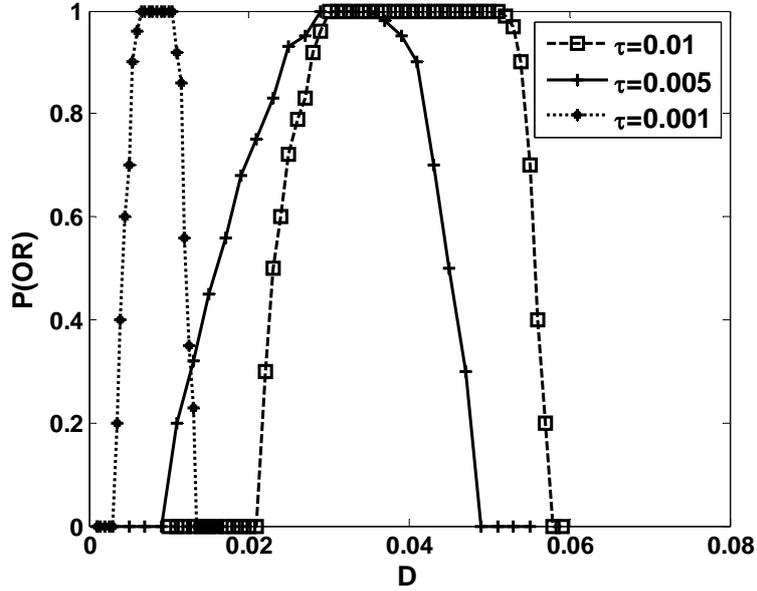

Fig.8. The success probability of obtaining the logical OR versus noise intensity $D$ with three different correlation time $\tau = 0.001$ (dotted-star line), $\tau = 0.005$ (solid–plus line), $\tau = 0.01$ (dash-rectangle line) for system (1) with parameters $a = 1$, $b = 2$, $x_l = -0.5$, $x_u = 1.5$, $A = 0$ and a fixed time step Dt=0.01.

Notably, we can get that the logic gate OR can be obtained at optimal noise intensity. Furthermore, with the increase of the correlation time $\tau$, the peak performance of the probability P(OR) moves to right and the optimal noise window becomes broader. About this we would not explain here because it is exactly the same with the explanation of Fig. 7.

We have known from Fig. 7, 8 that system (1) can realize the logic gate AND with thresholds $(x_l, x_u) = (-1.5, 0.5)$ and the logic gate OR with $(x_l, x_u) = (-0.5, 1.5)$ for an optimal noise value. Furthermore, with the purpose of ascertaining the specific range of the upper and lower thresholds in response to obtain the two different logic gates for an optimal noise value D, here we take D=0.05 and display the density maps of success probability P(AND) and P(OR) in Fig. 9.

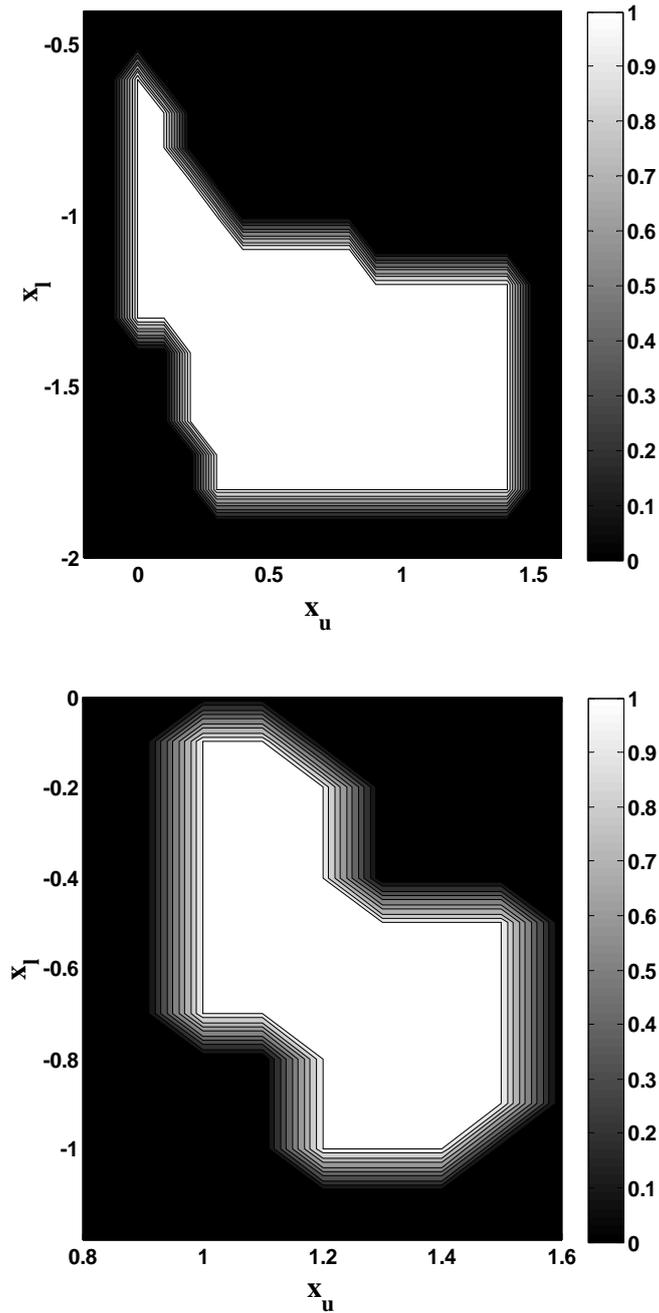

Fig.9. (Color on-line) From top to bottom: the first surface plot shows success

probability $P(\text{AND})$ and the second one shows $P(\text{OR})$. The x-axis displays the upper threshold $x_u$ and the y-axis displays the lower threshold $x_l$. With fixed parameters $a=1$, $b=2$, $D=0.05$, $\tau=0.01$, $Dt=0.01$. The light portion of the two plots indicate the logic gates AND and OR are obtained reliably.

As shown in Fig. 9, two different kinds of logic gates are obtained in different ranges of the upper and lower thresholds. This is owing to that the upper and lower thresholds determine the depths and position of the potential wells, and the thresholds determine the well the system falls in corresponding to the input stream. Therefore, with a suitable noise level, the logic gate AND can be switched into OR by simply adjusting the thresholds, which is a main feature in the dynamics.

Further, the logic gates AND and OR can also be obtained under optimal noise intensity in system (1) induced by Gaussian white noise (GWN) [10] instead of the dichotomous noise, and we show the density map of success probability for logic gate AND and OR in Fig. 10. From this figure we can observe that the best thresholds values for the desired logic gate can be found. In particular, it should be noted that the best ranges of the two thresholds for the logical AND or OR under GWN are smaller than dichotomous noise, which demonstrates that the dichotomous noise is superior to GWN in improving the availability of obtaining correct logic gate.

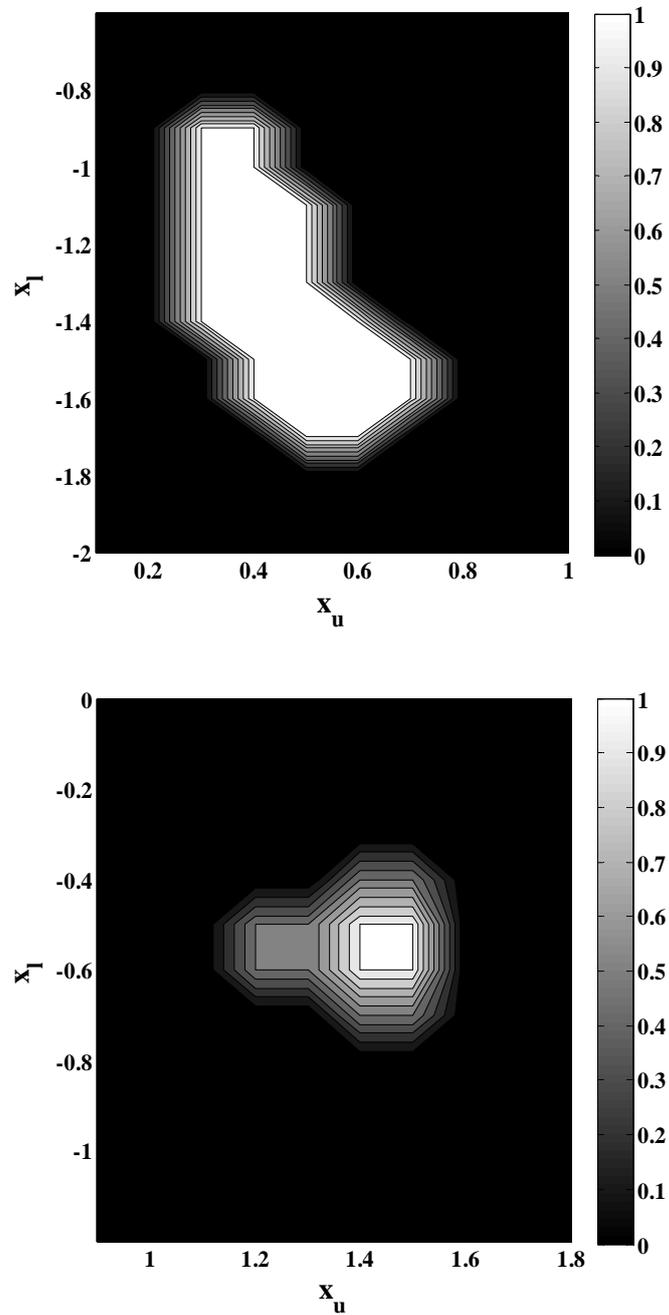

Fig.10. (Color on-line) From top to bottom: the first surface plot shows success probability $P(\text{AND})$ and the second one shows $P(\text{OR})$. The x-axis displays the upper threshold $x_u$ and the y-axis displays the lower threshold $x_l$. With fixed parameters in system (1) induced by Gaussian white noise: $a=1$, $b=2$,

$D = 0.5$, $Dt = 0.01$. The light portion of the two plots indicate the logic gates AND and OR are obtained reliably.

## 4. CONCLUSIONS

In summary, the LSR can be achieved in the nonlinear bistable system (1) induced by dichotomous noise. In this work, the bistable system driven by dichotomous noise displayed a phenomenon different from Gaussian white noise, which has been explained by observing effects of the logic inputs on the escape rate. Besides, the logic gate OR is gained when the inputs $(1,1)$ and $(0,1)/(1,0)$ sends the system to the right well and $(0,0)$ leads to the system in the left well; and the logic gate AND is acquired when the inputs $(0,0)$ and $(0,1)/(1,0)$ corresponding to the system states in the left well, and $(1,1)$ leads to the system in the right well. We investigated in the system (1) with thresholds $(x_l, x_u) = (-1.5, 0.5)$ with fixed noise asymmetry or with fixed noise correlation time or fixed noise intensity and obtained the logical AND. Moreover, for system (1) with thresholds $(x_l, x_u) = (-0.5, 1.5)$, it can realize the logic operation OR as we fix the noise correlation time. Namely, the logic operation can be switched into another by changing the thresholds of the system. And we also ascertain the specific ranges of the thresholds in obtaining the logic gates AND and OR for a suitable noise level. Furthermore, the influences of the noise asymmetry, noise correlation time and the noise intensity on the success probability have also been demonstrated respectively, which indicated an improvement in the reliability of achieving logic gate by optimizing parameters of dichotomous noise. And a fact is found that the dichotomous noise is superior to Gaussian white noise in enhancing the reliability of realizing logic operation.

ACKNOWLEDGEMENTS

This work was supported by the NSF of China (Grant Nos. 10972181, 11102157), Program for NCET, the Shaanxi Project for Young New Star in Science & Technology, NPU Foundation for Fundamental Research and SRF for ROCS, SEM.

REFERENCES

[1] Benzi, R., Sutera, A., Vulpiani, A.: The mechanism of stochastic resonance. J. Phys. A: Math. Gen. 14, L453-L457 (1981)

[2] Gammaitoni, L., Hänggi, P., Jung, P., Marchesoni, F.: Stochastic resonance. Rev. Mod. Phys. 70, 0034-6861 (1998)

[3] Bayley, H., Cremer, P. S.: Stochastic sensors inspired by biology. Nature 413, 226-230 (2001)

[4] Guderian, A., Dechert, G., Zeyer, K. P., Schneider, F. W.: Stochastic resonance in Chemistry.1.the Belousov-Zhabotinsky reaction. J. Phys. Chem. 100, 4437-4441 (1996)

[5] Gammaitoni, L., Marchesoni, F., Santucci, S.: Stochastic resonance as a Bona Fide Resonance. Phys. Rev. Lett. 74, 1052-1055 (1995)

[6] Harmer, G. P., Davis, B. R., Abbott, D.: A review of stochastic resonance circuits and measurement.

Transactions on Instrumentation and Measurement 51, 299-309 (2002)

[7] Zhang, L.Y., Cao, L., Wu, D. J., Wang, J.: Stochastic resonance in linear regime of a single-model laser. Chin. Phys. Lett. 20, 25-27 (2003)

[8] Merali, K., Rajamohamed, I., Sinha, S., Ditto, W. L., Bulsara, A. R.: Realization of reliable and flexible logic gates using noisy nonlinear circuits. Appl. Phys. Lett. 95, 194102_1-194102_3 (2009)

[9] Singh, K. P., Sinha, S.: Enhancement of "logical" responses by noise in a bistable optical system. Phys. Rev. E. 83, 046219_1-046219_6 (2011)

[10] Murali, K., Sinha, S., Ditto, W. L.: Reliable logic circuit elements that exploit nonlinearity in the presence of a noise floor. Phys. Rev. Lett. 102, 104101_1-104101_4 (2009)

[11] Zhang, L., Song, A. G., He, J.: Logical signals driven stochastic resonance in bistable dynamics subjected to 1/f noise floor. Eur. Phys. J. B 80, 147-153 (2011)

[12] Gupta, A., Sohane, A., Kohar, V., Murali, K., Sinha,S.: Noise-free logical stochastic resonance. Phys. Rev. E. 84, 055201_1-055201_5 (2011)

[13] Dari, A., Kia, B., Bulsara, A. R., Ditto, W. L.: Logical stochastic resonance with correlated internal and external noises in a synthetic biological logic block. Chaos 21, 047521_1-047521_8 (2011)

[14] Almog, R., Zaitsev, S., Shtempluck, O., Buks, E.: Signal amplification in a nanomechanical Duffing resonator via stochastic resonance. Appl. Phys. Lett. 90, 013508_1-013508_3 (2007)

[15] Zamora-Munt, J., Masoller, C.: Numerical implementation of a VCSEL-based stochastic logic gate via polarization bistability. Optics Express 18, 418-429 (2010)

[16] Guo, F., Zhou, Y. R.: Stochastic resonance in a stochastic bistable system subject to additive white noise and dichotomous noise. Physica A. 388, 3371-3376 (2009)

[17] Ghosh, P. K., Marchesoni, F., Savel'ev, S. E., Nori, F.: Geometric stochastic resonance. Phys. Rev. Lett. 104, 020601_1-020601_4 (2010)

[18] Ghosh, P. K., Glavery, R., Marchesoni, F., Savel'ev, S. E., Nori, F.: Geometric stochastic resonance in a double cavity. Phys. Rev. E 84, 011109_1-011109_7 (2011)

[19] Burada, P. S., Schmid, G., Reguera, D., Rubí, J. M., Hänggi, P.: Biased diffusion in confined media: Test of the Fick-Jacobs approximation and validity criteria. Phys. Rev. E 75, 051111_1-051111_20 (2007)

[20] Barik, D., Ghosh, P. K., Ray, D. S.: Langevin dynamics with dichotomous noise; direct simulation and applications. J. Stat. Mech. 3, 03010_1-03010_14 (2006)

[21] Kim, C., Lee, E. K.: Numerical method for solving differential equations with dichotomous noise. Phys. Rev. E 73, 026101_1-026101_11 (2006)

[22] Bena, I.: Dichotomous Markov noise: Exact results for out-of-equilibrium systems. Int. J. Mod. Phys. B 20, 2825-2834 (2006)

[23] Xu, Y., Wu, J., Zhang, H.Q., Ma, S.J.: Stochastic resonance phenomenon in an underdamped bistable system driven by weak asymmetric dichotomous noise. Nonlinear Dynamics 70, 531-539 (2010)

[24] Van den Broeck, C.: On the relation between white shot noise, Gaussian white noise, and the dichotomous Markov process. J. Stat. Phys. 31, 467-483 (1983)

[25] Fulinski, A.: Non-markovian dichotomous noise. Acta Physica Polonica B 26, 1131-1157 (1995)

[26] Ralls, K.S., Buhrman, R.A.: Defect interactions and noise in metallic nanoconstrictions. Phys. Rev. Lett. 60, 2434-2437 (1988)

[27] Pogany, D., Chroboczek, J.A., Ghibaudo, G.: Random telegraph signal noise mechanisms in reverse base current of hot carrier-degraded submicron bipolar transistors: Effect of carrier trapping during stress on noise characteristics. J. Appl. Phys. 89, 4049-4058 (2001)

[28] Xi, H.W., Loven, J., Netzer, R., Guzman, J.I., Franzen, S., Mao, S.: Thermal fluctuation of magnetization and random telegraph noise in magnetoresistive nanostructures. J. Phys. D: Appl. Phys. 39, 2024-2029 (2006)

[29] Rogers, C.T., Buhrman, R.A.: Composition of 1/f noise in metal-insulator-metal tunnel junctions. Phys. Rev. Lett. 53, 1272-1275 (1984)

[30] Hu, G.: Stochastic forces and nonlinear systems, Shanghai Scientific and Technological Education Publish House. SHANGHAI, 221 (1994).

[31] Schmid, G. J., Reimann, P., Hänggi, P.: Control of reaction rate by asymmetric two-state noise. J. Chem. Phys. 111, 3349-3356 (1999)

[32] Bulsara, A.R., Dari, A., Ditto, W.L., Murali, K., Sinha, S.: Logical stochastic resonance. Chem. Phys. 375, 424-434 (2010)